\documentclass[english,prl,twocolumn]{revtex4}
\usepackage[latin9]{inputenc}
\usepackage{amsmath}
\usepackage{graphicx}
\usepackage{amssymb}
\newcommand{\ket}[1]{| #1 \rangle}

\begin{document}

\title{Superconducting resonators
as beam splitters for linear-optics quantum computation}

\author{Luca Chirolli}
\author{Guido Burkard}

\affiliation{Department of Physics, 
University of Konstanz, D-78457 Konstanz, Germany}

\author{Shwetank Kumar}
\author{David P. DiVincenzo}

\affiliation{IBM T. J. Watson Research Center, Yorktown Heights, NY 10598 USA}


%


\maketitle

{\bf
A functioning quantum computer will be a machine that builds up, in a programmable way, nonclassical correlations in a multipartite quantum system. Linear optics quantum computation 
(LOQC)  \cite{KLM,LOQCrev} is an approach for achieving this function
that requires only simple, reliable linear optical elements, namely beam splitters and phase shifters.  Nonlinear optics is only required in the form of single-photon sources for state initialization, and detectors.  However, the latter remain difficult to achieve with high fidelity.
A new setting for quantum optics has arisen in circuit quantum electrodynamics (cQED) using superconducting (SC) quantum devices \cite{Wallraff}, and opening up the way to LOQC using microwave, rather than visible photons.  Much progress is being made in SC qubits \cite{DiCarlo} and cQED:  high-fidelity Fock state generation \cite{HofheinzNature09} and qubit measurements provide single photon sources and detection.  Here we show that the LOQC toolkit in cQED can be completed with high-fidelity ($>99.92\%$) linear optical elements.
}

We propose and analyze a technique for producing a beam-splitting quantum gate between two modes of a ring-resonator SC cavity.  The cavity has
two integrated superconducting quantum interference devices (SQUIDs) that are modulated by applying an external magnetic field.  The gate is accomplished by applying a radio frequency pulse to one of the SQUIDs at the difference of the two mode frequencies. Departures from perfect beam splitting only arise from corrections to the rotating wave approximation; an exact calculation gives a fidelity of $>0.9992$.

We analyze in detail here a realization of a beam splitter; the phase shifter represents a simpler case that
can be studied in exactly the same way.  Of course, a beam splitter is a well known device in microwave technology, where it is known as the directional coupler.  But we need to achieve the functioning of a beam splitter in a different way, because the microwave photons in cQED are never ``flying''; we must find a way to accomplish the action of beam splitting in place between photonic modes that remain resident in SC resonators.  

In fact we show here that high-fidelity beam splitting can be produced in exactly the ring-resonator device analyzed by Ref.~\cite{KumarDiVincenzo09} (see Fig. 1a).  This structure has two nearly degenerate fundamental modes, even and odd with respect to the horizontal midline, with frequencies $\omega_{1,2}$ respectively.  The beam-splitting action will take place between photons in these two modes.  This ring is interrupted by two SQUIDs as shown. In \cite{KumarDiVincenzo09} it is shown that the nonlinearity of these SQUIDs is enough to enable nondemolition measurements; but this nonlinearity is far too small to make a practical traditional quantum gate such as a cPHASE.  We can neglect this nonlinearity here, and we will exploit another control available in this device: the effective inductance of these SQUIDs can be controlled by the external magnetic fluxes shown. This inductance couples to the photonic modes by setting the reflection coefficient for standing waves around the ring.  It is by {\em modulation} of the inductance of SQUID$_1$ at the difference frequency $\omega_1-\omega_2$ that high-fidelity beam splitting action is achieved  \cite{JRM}.

We set up an accurate analysis of the photonic dynamics of our ring resonator by representing the rest of the ring connected to SQUID$_1$ as a linear, passive structure with two-terminal impedance $Z(\omega)$.  
Note that we will consider the flux bias of SQUID$_2$ to be fixed, so that it functions as an inductor with, using the parameters of \cite{KumarDiVincenzo09}, $L_0=0.19$ nH. We neglect losses so that $Z$ is purely imaginary.  For device parameters from \cite{KumarDiVincenzo09}, $Z(\omega)$ can be calculated analytically using standard two-port theory\cite{ABCD}. ${\rm Im} Z(\omega)$ in the frequency range of interest is shown in Fig. 1c. Since we include in $Z$ the inductance $L_0$ representing the average of the modulated SQUID$_1$, the two zeros of $Z(\omega)$ correspond to the two mode frequencies $\omega_{1,2}$ of the unmodulated ring resonator.

\begin{figure}[t]
\begin{center}
 \includegraphics[width=8cm]{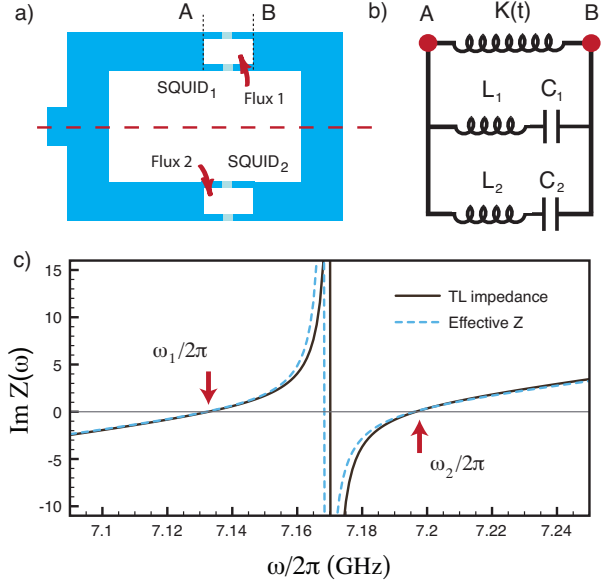}
   \caption{ a) Schematic representation of the superconducting ring resonator, comprising two SC transmission line segments coupled by two SQUIDs that can be externally controlled by two different applied magnetic fluxes.  AC modulation of SQUID$_1$ will accomplish the action of beam splitting between the two fundamental modes of the ring resonator. The device can optionally be tuned by a stub on the midline as shown. b) Equivalent circuit, consisting of two parametrically coupled LC resonators.  The modulation of the SQUID$_1$ impedance is represented by the time-dependent inductor $K(t)$.  The LC circuit parameters are chosen so that the impedance of the two parallel resonators matches the two-terminal impedance, looking down into the AB port, of the ring resonator, in a frequency band including the two fundamental modes.  c) The AB-port impedance of the resonator and of the equivalent circuit. The unmodulated fundamental mode frequencies $\omega_{1,2}$ are given by the two zeros of this impedance.   \label{Fig1}}
\end{center}
\end{figure}

This impedance is extremely well reproduced (dashed line) by that of the two-pole structure in our equivalent circuit shown in Fig.~1b.  This fit is obtained with parameters $L_1=5.8$ nH, $C_1=86$ fF, $L_2=7.7$ nH, and $C_2=63$ fF.  It is practical for the amplitude of the parametrically modulated inductance $K(t)$ to be around 0.04 nH.  We can apply our network graph theory \cite{BKD} to analyze the quantum behavior of these modes of the ring resonator. The classical time-dependent Hamiltonian that describes the equivalent circuit Fig.~\ref{Fig1}b is
\begin{equation}
H=\sum_{i=1,2}\left(\frac{Q_i^2}{2C_i}+\frac{\Phi_i^2}{2L_i}\right)
-\frac{K(t)}{2}\left(\frac{\Phi_1}{L_1}+\frac{\Phi_2}{L_2}\right)^2.
\label{eq:circuit-Hamiltonian}
\end{equation}
We choose the time dependent inductance to have the form $K(t)=\delta K \cos(\omega_d t)$.  We will present calculations only for the case where $\delta K$ is switched on from zero at $t=0$, then switched off again at $t=\tau$.  We will see that even for such an unshaped, square modulation pulse, the desired quantum gate operation can be achieved with excellent fidelity.  We find that this fidelity is insensitive to details of pulse shaping.

We quantize Eq. (\ref{eq:circuit-Hamiltonian}) by imposing commutation rules between canonically conjugate variables $[\hat{\Phi}_i,\hat{Q}_j]=i\hbar\delta_{ij}$ and express them, using
creation and annihilation operators $a^\dag$ and $a$, as
$\hat{\Phi}_i=\sigma_i\sqrt{\hbar}(a_i+a^{\dag}_i)/\sqrt{2}$, and 
$\hat{Q}_i=-i\sqrt{\hbar}(a_i-a^{\dag}_i)/\sqrt{2}\sigma_i$, 
with $\sigma_i=(L_i/C_i)^{1/4}$. 
Assuming $K(t)\ll L_{1,2}$, the Hamiltonian becomes ($\hbar=1$)
\begin{equation}\label{Eq1:Ham}
H=\sum_{i=1,2}\omega_i a_i^{\dag}a_i+f(t)\left[\lambda(a_1+a_1^{\dag})+\frac{1}{\lambda}(a_2+a_2^{\dag})\right]^2.
\end{equation}
Here the two resonant harmonic frequencies are $\omega_i=1/\sqrt{L_iC_i}$, $\lambda=(L_2^3C_2/L_1^3C_1)^{1/8}$, and  $f(t)=-\sigma_1\sigma_2K(t)/4L_1L_2\equiv f\cos(\omega_dt)$.
We will consider driving at resonance at transition frequency $\omega_d=\Delta\omega=\omega_1-\omega_2$.  For the ring resonator we will always be in the regime $\Delta\omega\ll\omega_1,\omega_2$.

To study the beam splitting action created by the $K(t)$ pulse, we will calculate the time evolution of an initial state comprising a single photon in one of the two modes (mode 1), i.e., $|\psi_i\rangle=a_1^{\dag}|0\rangle$, with $|0\rangle=|0\rangle_1|0\rangle_2$.
We aim for the final ``beam-split'' state $|\psi_f\rangle=\frac{1}{\sqrt 2}(a_1^{\dag}-ia_2^{\dag})|0\rangle$.  We consider always a doubly-rotating frame with rotation at frequency $\omega_{1,2}$ for modes 1,2 resp.  We will study a fidelity ${\cal F}(\tau)=|\langle \psi_f|{\cal U}(\tau)|\psi_i\rangle|^2$ which indicates how close to ideal beam-splitting our operation is.
Here, ${\cal U}(t)={\cal T}\exp (-i\int^tdt'H(t'))$ denotes the evolution operator generated by the Hamiltonian Eq.~(\ref{Eq1:Ham}).  The use of more general gate fidelities would not affect our conclusions.

We present three approaches to the calculation of ${\cal F}(\tau)$.  First we perform a naive rotating wave approximation (RWA) in which only time independent terms in the rotating frame are retained in $H(t)$.  An elementary calculation gives ${\cal F}_{\rm RWA}(\tau)=(1+\sin(2 f \tau))/2$.  Given the smallness of $\Delta\omega$, a second calculation is much more accurate, which retains additional terms in the rotating frame that oscillate at frequencies $\Delta\omega$ and $2\Delta\omega$. Then the part of the Hamiltonian that generates the beam-splitting gate is 
\begin{equation}
H_{\rm BS}=f~a_1a^{\dag}_2(1+e^{-2i\Delta \omega t})+{\rm h. c.},
\end{equation}
The terms oscillating at frequency $2\Delta \omega$ produce Bloch-Siegert oscillations (BSO) \cite{BlochSiegert}. For $f\ll\Delta\omega$ these can be treated perturbatively \cite{ShahriarPRA04}. By considering virtual transitions via the first two sidebands shifted in energy by $\pm 2\Delta\omega$, the fidelity at first order in $f/\Delta\omega$  is
\begin{equation}\label{Eq:FidelityRWA}
{\cal F}_{\rm BSO}(\tau)=\frac{1}{2}\left[1+\sin(2f \tau)+\frac{f}{\Delta\omega}\cos(2f \tau)\sin(2\Delta\omega \tau)\right].
\end{equation}
Within both the RWA and the BSO approximations, for a pulse of duration $\tau=\pi/4 f$ we attain ${\cal F}=1$: beam splitting is perfect at these levels of approximation.  There is a real difference between these two points of view; in the naive RWA the Bloch vector in the $\ket{01}$-$\ket{10}$ space undergoes simple circular motion, while the BSO manifest themselves as a nutational motion of the Bloch vector (Fig. 2).  But in both cases the Bloch vector arrives exactly at the $x$-axis. 
Note that in both cases the Rabi oscillation frequency $\Omega_R=2 f$ can be simply expressed as 
\begin{equation}
\Omega_R=\frac{\delta K}{2}\sqrt{\omega_1\omega_2\over L_1L_2}.
\end{equation}
For the parameters of our ring resonator, this gives a very convenient value of $\Omega_R/2\pi\approx 20$MHz.

\begin{figure}[t]
\begin{center}
 \includegraphics[width=8cm]{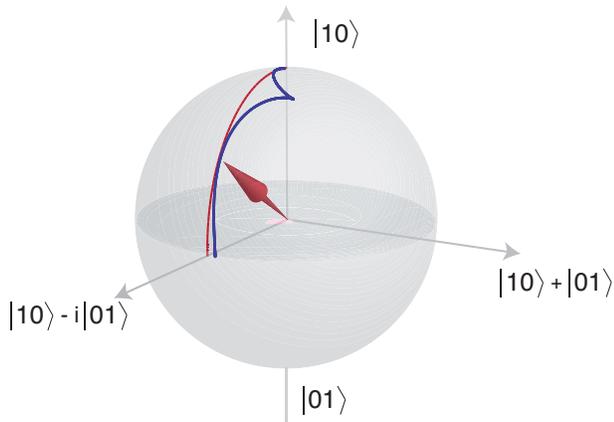}
   \caption{To good approximation, the quantum evolution is confined to the Bloch sphere defined by the $\ket{01}$ and $\ket{10}$ states. Counterrotating terms in the Hamiltonian at frequency $2\Delta\omega$ perturb the simple Bloch-sphere (red) evolution by terms of order $f/\Delta\omega$. Bloch-Siegert oscillations (blue) cause nutation of the Bloch vector superimposed on regular precession in the rotating frame. The device parameters for the ring resonator are as given in the text. \label{Fig2}}
\end{center}
\end{figure}

Unfortunately, the actual time evolution of our Hamiltonian does not give 100\% fidelity for beam splitting.  Parametric time-dependent modulation causes mode squeezing; in other words, photon number is not conserved, and our evolution does not remain confined to the $\ket{01}$-$\ket{10}$ Bloch sphere. To quantify this effect, we must do a third calculation that goes beyond any rotating wave approximation.  With only a modest amount of numerical effort, it is feasible to do an essentially exact calculation of our gate operation.  All that is required is a $4\times 4$ matrix calculation of the Heisenberg operators $a_{1,2}(t)$ and $a^\dagger_{1,2}(t)$\cite{Scully}.  This calculation begins by using the canonically conjugate quadratures $\boldsymbol{\xi}=(\hat{q}_1,\hat{q}_2,\hat{p}_1,\hat{p}_2)^T$ that are related to the original fluxes and charges by $(\hat{\Phi}_1,\hat{\Phi}_2,\hat{Q}_1,\hat{Q}_2)^T=\sqrt{\hbar}D_{\sigma}\boldsymbol{\xi}$, with the diagonal matrix $D_{\sigma}={\rm diag}(\sigma_1,\sigma_2,1/\sigma_1,1/\sigma_2)$. 
Because the Hamiltonian governing the evolution is quadratic, the Heisenberg equations of motion for the canonical quadratures $\boldsymbol{\xi}$ are linear,
\begin{equation}
\dot{\boldsymbol{\xi}}=\Xi(t)\boldsymbol{\xi}
=\left(\begin{array}{cc}
0 & \Omega\\
-\Omega-4f(t)\Lambda & 0
\end{array}\right)
\boldsymbol{\xi},
\end{equation}
where $\Lambda$ and $\Omega$ are real $2\times 2$ matrices with $\Omega={\rm diag}(\omega_1,\omega_2)$, $\Lambda_{11}=\lambda^2$, $\Lambda_{22}=1/\lambda^2$ and $\Lambda_{12}=\Lambda_{21}=1$. 
The general solution can be expressed in terms of $S(t)={\cal T}\exp\int^tdt'\Xi(t')$, where $S(T)$ is a $4\times 4$ real symplectic matrix that satisfies  $S^T(t)JS(t)=J$, with the $4\times 4$ real antisymmetric matrix $J$ having a $2\times 2$ block structure, with $J_{12}=-J_{21}=\openone$ and $J_{11}=J_{22}=0$. 

The action of the evolution operator ${\cal U}(t)$ on the canonical quadratures $\boldsymbol{\xi}$ results in a matrix multiplication $\boldsymbol{\xi}(t)={\cal U}^{\dag}(t)\boldsymbol{\xi}{\cal U}(t)=S(t)\boldsymbol{\xi}$ that connects the Heisenberg and the Schr\"odinger representation. A similar relation holds for the field operators $a_i$ and $a^{\dag}_i$. By defining the vector ${\bf a}=(a_1,a_2,a^{\dag}_1,a^{\dag}_2)^T$, the connection between the Heisenberg and the Schr\"odinger representation reads as ${\bf a}(t)=S_{(c)}(t){\bf a}$, with $S_{(c)}(t)=\Sigma^{\dag}S(t)\Sigma$, where $\Sigma$ is the simple unitary matrix of the basis change $\boldsymbol{\xi}=\Sigma{\bf a}$.

\begin{figure}[t]
\begin{center}
 \includegraphics[width=8.5cm]{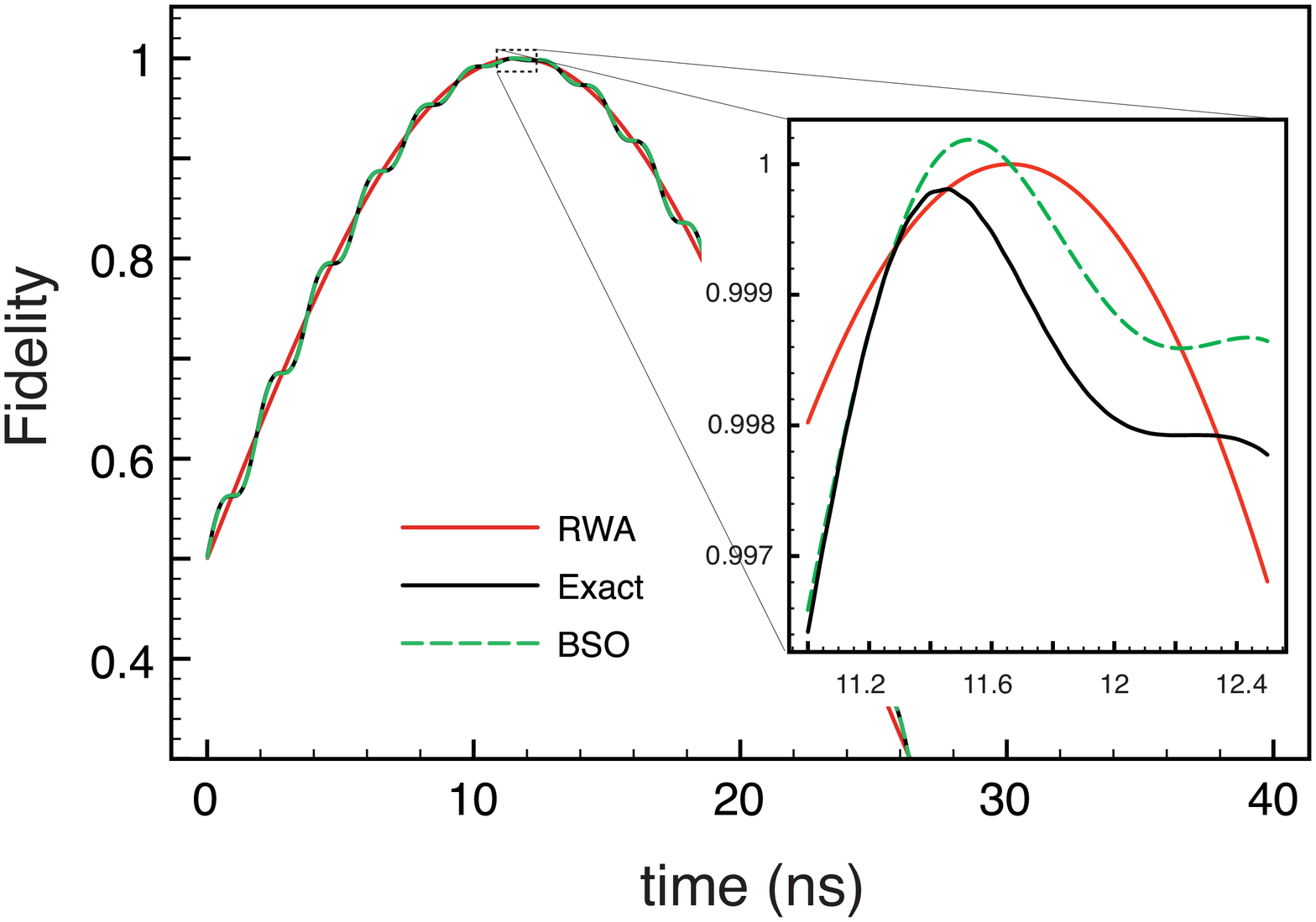}
   \caption{Fidelity of the beam-splitting gate as a function of pulse duration, using a naive rotating wave approximation (RWA), a rotating wave approximation including some additional slow-rotating terms perturbatively, capturing the occurrence of Bloch-Siegert oscillations (BSO), and with an essentially exact numerical calculation.  Due to its simple perturbative nature, the BSO fidelity can (and does) exceed 1.  The fidelity can 
be further optimized by carefully adjusting the
resonator frequencies and thus the phase of the BSO.  The device parameters for the ring resonator are as given in the text, except that stub tuning is used to raise the mode splitting to $\Delta\omega\approx 260$~MHz. \label{Fig3}}
\end{center}
\end{figure}

Here we show that the real symplectic matrix $S(t)$ contains all the information needed to calculate the fidelity ${\cal F}(\tau)=|\langle\psi_f|{\cal U}(t)|\psi_i\rangle|^2$ exactly. Any real symplectic matrix $S\in~{\rm Sp}(4,\mathbb{R})$ admits a singular value decomposition in terms of real orthogonal symplectic matrices, $S=S_LDS_R^T$, with $S^T_{\alpha}S_{\alpha}=\openone$, $S_\alpha^TJS_{\alpha}=J$, for $\alpha=L,R$, and $D={\rm diag}(\kappa_1,\kappa_2,1/\kappa_1,1/\kappa_2)$, which is unique up to a reordering of the diagonal entries of $D$ \cite{Xu}.  This decomposition of $S$ induces a corresponding decomposition of the evolution operator ${\cal U}(S)$: ${\cal U}(S)={\cal U}(S_L){\cal U}(D){\cal U}^{\dag}(S_R)$. 
The elements $S_{\alpha}$ ($\alpha=L,R$) have the general $2\times 2$ block form $[S_{\alpha}]_{11}=[S_{\alpha}]_{22}=X_{\alpha}$ and $[S_{\alpha}]_{12}=-[S_{\alpha}]_{21}=Y_{\alpha}$, 
with $X_{\alpha}, Y_{\alpha}$ real $2\times 2$ matrices such that $U_{\alpha}\equiv X_{\alpha}-iY_{\alpha}$ is a unitary $2\times 2$ matrix \cite{Note,Arvind94}. 

It follows that the unitary evolution ${\cal U}(S_{\alpha})$ associated with $S_{\alpha}$ conserves the number of photons and does not mix the creation and annihilation operators: ${\cal U}^{\dag}(S_{\alpha})a_i{\cal U}(S_{\alpha})=\sum_{k=1,2}[U_{\alpha}]_{jk}a_k$  \cite{Arvind95}. On the other hand, the diagonal matrix $D$ represents an active term that introduces squeezing in the two modes. In terms of mode operators ${\cal U}(D)$ can be expressed using independent squeezing operators ${\Pi}_i(\rho)=\exp[\rho(a_i^2-a_i^{\dag}{}^2)/2]$: ${\cal U}(D)={\cal U}(\kappa_1,\kappa_2)={\Pi}_1(-\ln\kappa_1){\Pi}_2(-\ln\kappa_2)$.  Its action on the mode annihilation operators is simply given by ${\Pi}_i^{\dag}(\rho)a_i{\Pi}_i(\rho)=a_i \cosh\rho-a_i^{\dag}\sinh\rho$. These facts allow us to write the fidelity as 
\begin{equation}
{\cal F}=|\langle\Psi_f|{\cal U}(\kappa_1,\kappa_2)|0\rangle|^2,\label{fid3}
\end{equation}
with the state $|\Psi_f\rangle=(b_1^++b_2^+)|00\rangle/\sqrt{2}+(b_1^-+b_2^-)|11\rangle/\sqrt{2}-b_1^-|20\rangle-b_2^-|02\rangle$ with $b_j^{\pm}=\delta_j(\kappa_j^2\pm1)/2\kappa_j$ and $\delta_j=[U_R]_{1j}([U_L]_{1j}+i[U_L]_{2j})^*$ for $j=1,2$. Eq. (\ref{fid3}) can now be evaluated in closed form, because the two-mode squeezed vacuum ${\cal U}(\kappa_1,\kappa_2)|0\rangle$ has an analytical expression in the Fock space via the known relation \cite{Scully} 
\begin{equation}
\langle n|{\Pi}(\rho)|0\rangle=\frac{(\tanh \rho)^{n/2}}{2^{n/2}(n!\cosh\rho)^{1/2}}H_n(0),
\end{equation}
with $H_n(0)$ the Hermite polynomial at the origin, which is zero for odd $n$ and $H_n(0)=(-1)^{n/2}n!/(n/2)!$ for even $n$. Therefore, after a singular value deomposition of $S$ the fidelity is directly obtained as a simple function of the quantities $\kappa_i$ and $\delta_i$.

Fig. 3 shows our three calculations of the beam-splitting fidelity as a function of pulse duration time $\tau$.  The device parameters for the ring resonator are as given above, except that by introducing an electrical stub for tuning as in Fig. 1a, the frequency of the even-symmetry mode $\omega_1/2\pi$ is lowered to around 6.93 GHz, so that $\Delta\omega/2\pi\approx 260$ MHz.  We see that the evolution approximately follows the smooth Rabi oscillation predicted by the naive RWA, but that there are appreciable Bloch-Siegert oscillations superimposed on this. The perturbative BSO calculation in fact comes very close to the exact evolution for our parameters.  The exact caclulation gives an extremely high value of the fidelity: ${\cal F}_{max}>0.9992$. Squeezing is very small because the parametric modulation frequency is very slow compared with the mode frequencies ($\Delta\omega\ll\omega_{1,2}$).  For the ring resonator without the stub this ratio is even smaller, since then $\Delta\omega\approx 64$ MHz; but this device is awkward to use since then the Bloch Siegert oscillations become very large and are no longer well described perturbatively \cite{Fuchs}.

To summarize, we see that for a very straightforward ring-resonator geometry, almost ideal beam-splitting between two cavity modes can be readily achieved.  The calculated fidelity of 0.9992 is not realistic; other limits such as resonator loss and 1/f noise would come into play at this level for the present state of the art.  But our result shows that there is no intrinsic limit to accomplishing effective beam splitting by device modulation.  Finally, we mention that an effective ``delay line" is also readily implemented in this device; by transiently changing the DC bias fluxes of the two SQUIDs, in either an even or an odd fashion, either mode may be subjected to any desired phase shift.  There will be even less intrinsic limitation on the fidelity of these operations.  Thus, we see that the toolkit for linear optics quantum computing is readily completable in superconducting microwave circuits. 

Financial support from SFB 767 ``Controlled Nanostructures'' and the Konstanz Center for Applied Photonics (CAP) are gratefully acknowledged. This
research was  supported in part by the National Science Foundation
under Grant No. PHY05-51164.


\begin{thebibliography} {99}
\bibitem{Note}
The elements $S_{\alpha}$ belong to the maximally compact subgroup K(2) of ${\rm Sp}(4,\mathbb{R})$ that is isomorphous to U(2), ${\rm K}(2)={\rm SO}(4)\cap{\rm Sp}(4,\mathbb{R})={\rm U}(2)$.
\end{thebibliography}
\end{document}